\documentclass[twocolumn]{emulateapj}
\usepackage{multirow}
\usepackage{graphics}
\usepackage{amsmath, amsthm, amssymb}
\usepackage{epsfig}
\usepackage[hidelinks]{hyperref}
\usepackage{natbib}
\usepackage{float}
\usepackage[title]{appendix}
\usepackage{lineno}

\usepackage{graphicx}
\usepackage[dvipsnames]{xcolor}

\begin{document}

\title{
Lindblad and Corotation Resonances as the Common Origin of Moving Groups, Diagonal Ridges, and Velocity Waves in the Milky Way Disc
}

\author{Tatiana A. Michtchenko$^{1}$}\email[e-mail: ]{tatiana.michtchenko@iag.usp.br}
\author{Willian Nacafucasaco$^{1}$}\email[e-mail: ]{willianyn@usp.br}
\author{Wilton S. Dias$^{2}$}\email[e-mail: ]{wilton@unifei.edu.br}
\author{Douglas A. Barros$^{3}$}\email[e-mail: ]{douglas.barros@alumni.usp.br}

\affiliation{$^1$IAG, Universidade de S\~ao Paulo, Rua do Mat\~ao, 1226, Cidade Universit\'aria, 05508-090 S\~ao Paulo, Brazil\\
$^2$Instituto de Física e Química, Universidade Federal de Itajubá, Av.\,BPS 1303 Pinheirinho, 37500-903 Itajubá, MG, Brazil  \\
$^3$Escola T\'ecnica Estadual Professor Agamenon Magalh\~aes, Avenida Jo\~ao de Barros, 1769, Espinheiro, 52021-180, Recife, Brazil}

\date{\today}

\begin{abstract}

The Galactic disc is a complex dynamical system showing various kinematic structures, including moving groups (MGs), diagonal ridges (DRs), and velocity waves. To investigate their possible dynamical connection, we analyzed the phase-space distribution of stars from the \textit{Gaia} DR3 catalog. 
We used angular momentum $L_z$ as a key dynamical tracer and examined stellar densities and velocities as functions of $L_z$. We identified stellar groups at specific $L_z$-values and associated them with extrema in the median radial velocity $\left<V_R\right>$, known as a velocity wave. By cross-matching the extrema of the $\left<V_R\right>$-wave with structures in the $V_R$--$L_z$ and $R$--$L_z$ planes, we established a clear connection between the velocity wave, the main MGs, and the DRs. A similar connection was partially identified for corrugations in the median vertical velocity $\left<V_z\right>$. Both velocity waves are strongest near the Galactic mid-plane and decrease in amplitude with increasing vertical distance. Our results demonstrate that MGs, DRs, and velocity waves in the solar neighborhood are different projections of the same dynamical structures associated with Lindblad and corotation resonances. These resonances arise from coupling between the radial and azimuthal stellar motions and are particularly prominent in the thin Galactic disc. Numerical simulations including perturbations from the Galactic bar and spiral arms confirm these conclusions. Our findings indicate that the local stellar disc is currently evolving in a state of dynamical equilibrium. We also show that in-plane velocity oscillations must be considered when interpreting vertical stellar distributions, as they may produce artifacts in vertical profiles.

\end{abstract}


\maketitle

\section{Introduction}\label{sec:intro}

The oscillatory structures detected in the Galactic disc were initially identified by \cite{2012ApJ...750L..41W}, who reported a North–South asymmetry in both the number density and the bulk motions of stars in the solar neighborhood. The investigation of asymmetry has been extended to higher values of $|Z|$ using the improved precision of the \textit{Gaia} DR2 data, together with APOGEE and GALAH (Bennett and Bovy 2019). Wave-like patterns in the mean vertical and radial velocities of stars, plotted as  functions of their guiding-center radius $R_G$, or equivalently, their angular momentum component $L_z$, were described in \cite{Schonrich2018MNRAS} and \cite{2019MNRAS.490.5414F}. 

When asymmetry in the distribution of the kinematic stellar properties is regarded as evidence of the incomplete phase-mixing process ongoing in the Galactic disc, its origin is commonly attributed to recent external perturbations, one of these a Sagittarius dwarf galaxy-like merger (\citealp{2018Natur.561..360A}; \citealp{2018MNRAS.481.1501B}; \citealp{2023MNRAS.521..114T}). Consistent with this scenario, \cite{HUNT2025NEWAR} reviewed several numerical simulations, which demonstrated that external perturbations are capable of qualitatively reproducing the wave-like patterns observed in stellar kinematics (e.g., \citealp{2021MNRAS.504.3168B}; \citealp{2022A&A...668A..61A}; \citealp{2022MNRAS.515.5951T}). 

In this work, we adopt a different perspective: We assume that the local stellar disc is in dynamical equilibrium and attribute the non-uniform spatial and kinematic distribution of stars observed in this region to the disturbing effects of the asymmetric Galactic structures. Indeed, stellar orbits evolve within the  gravitational potential, whose structure is intrinsically non-axisymmetric, with asymmetries arising mainly from the spiral arms and the central bar. These components generate a complex phase–space portrait characterized by chains of resonant islands and bifurcation layers, which are naturally reflected in the distribution of the kinematic properties of the stars (e.g., \citealp{Michtchenko2018ApJL}; \citealp{2026ApJ..1000..236N}). Understanding the effects induced by the gravitational potential associated with the adopted Hamiltonian model, we can expect that their imprints will be discernible in the observable properties of the stellar phase-space distribution.

Following our approach, we carried out an in-depth analysis of the stellar distribution profile and their kinematic characteristics. 
For this purpose, we used a sample of stars with complete phase-space data from the \textit{Gaia} DR3 catalog \citep{2023A&A...674A...1G}, and adopted the angular momentum component $L_z$ as a fundamental tracer of the stellar kinematic properties. 
By evaluating the median radial and vertical velocities as functions of the angular momentum $L_z$, we detected wave-like kinematic patterns. Subsequently, we superposed and compared the principal characteristics of these features with previously identified asymmetric structures in the phase space, specifically the moving groups (MGs) and the diagonal ridges (DRs). 

Our results reveal a quantitative agreement between the extrema of the $\left<V_R\right>$-oscillation and the nominal locations of the main MGs projected at the $L_z$ direction. The same groups known as DRs when projected along the Galactocentric distance $R$,  produce some corrugations of both  $\left<V_R\right>$ and $\left<V_z\right>$ as functions of $R$. The results were further corroborated by numerical simulations of test-particle motions in the Galactic gravitational potential, incorporating perturbations from both the central bar and the spiral arms.

\section{Sample {and stellar distributions}}\label{sec:sample}

Our sample contains stars from \textit{Gaia} DR3 catalog, with astrometry quality measures limited to $ruwe < 1.4$, ensuring solved values of astrometric parameters {(\citealp{2021A&A...649A...2L})}. Moreover, we choose all objects for which line-of-sight velocity measurements have an {absolute  uncertainty less than 5\,km\,s$^{-1}$}.  Finally, we restricted our sample to include only objects that have more than 8 transits, to enclose objects with a good amount of observation cycles, and excluded binary stars to avoid internal variations in the radial velocity.

Adopting the Sun's position at $R_\odot=8.122$\,kpc, $\varphi_\odot = 90$\,deg and $Z_\odot = 15$\,pc, and the velocity vector $\left(-12.9,245.6,7.78\right)$\,km\,s$^{-1}$ (\citealp{2018RNAAS...2..210D}), we calculate the cylindrical coordinates  $\left(R,\varphi,Z\right)$ and velocity components $\left(V_R,V_\varphi,V_z\right)$ of the objects in the Galactocentric reference frame. 

We separate the sample into three sub-samples, as shown below:\\~

- sample\,\textbf{\#1} includes the stars belonging to the thin Galactic disc\footnote{According to \cite{2016ARA&A..54..529B}, the vertical size scale and the star dispersion velocities of the thin disc are within the ranges of $220$--$450$\,pc and $20$--$27$\,km\,s$^{-1}$, respectively.} defined as $|Z| < 200$\,pc and $|V_z| < 20$\,km\,s$^{-1}$ (total $\sim$ 5.8 million stars);\\~  

- sample\,\textbf{\#2} includes the stars that are located out of the Galactic thin disc, that is, the stars from the thick Galactic disc\footnote{The scale-height associated with the thick disc is around 650\,pc to 1000\,pc, approximately (\citealp{2002ApJ...578..151S}, \citealp{2018MNRAS.479..211M} and \citealp{2023Galax..11...77V}). For instance, it is reasonable to assume a wider region for stars belonging to the thick disc because of the uncertainty on those measurements.} defined in the  range of $|Z|$ between 500\,pc and 2\,kpc (total $\sim$ 5.7 million stars);\\~

- sample\,\textbf{\#3} includes the stars that are located at $|90^\circ-\varphi| < 6^\circ$ from the Sun–Galactic Center line (total $\sim$ 9.8 million stars). This condition is similar to the one used in \cite{HUNT2025NEWAR}.\\~

\begin{figure}
    \includegraphics[width=1.\columnwidth]{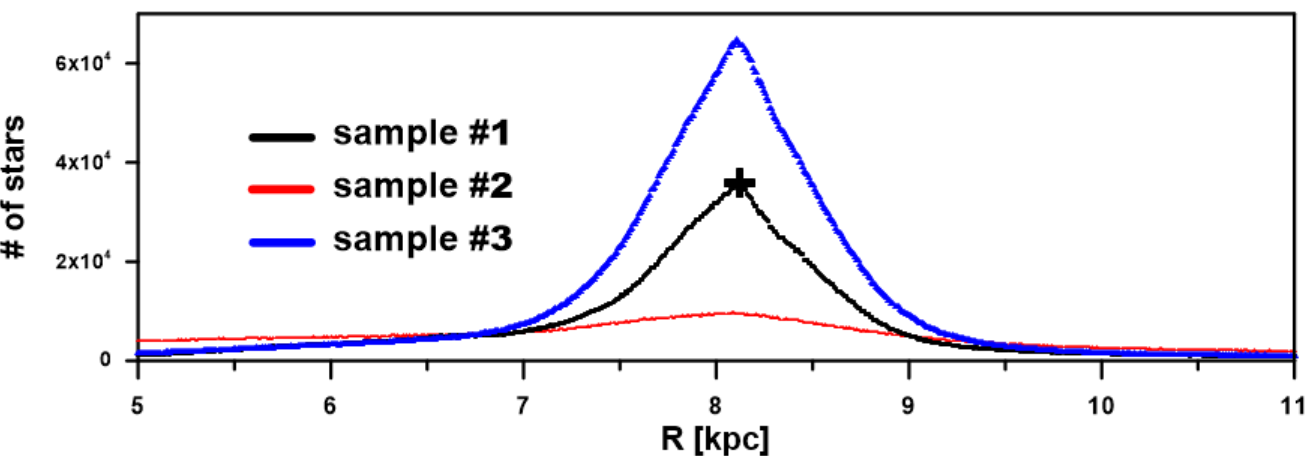}
    \includegraphics[width=1.\columnwidth]{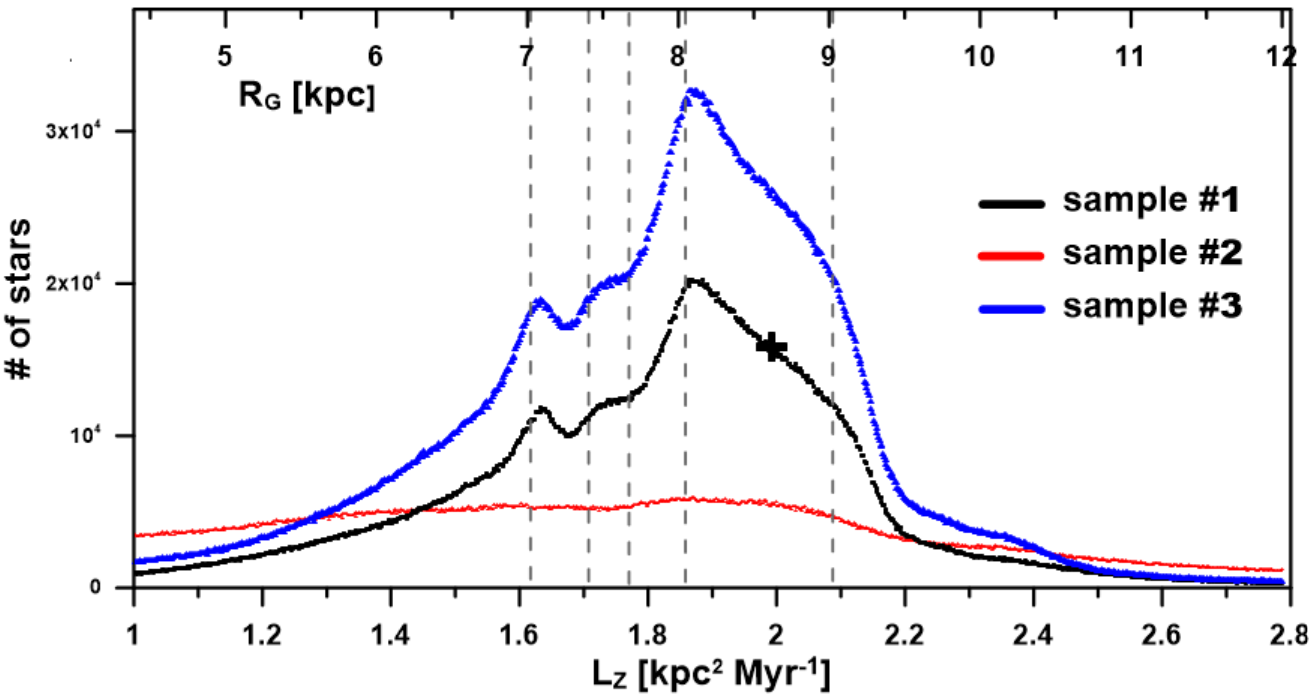}
    \caption{Top: Distributions of stars from the samples \textbf{\#1}, \textbf{\#2} and \textbf{\#3} as functions the Galactocentric distance $R$. The position of the Sun is shown by a cross symbol.  
    Bottom: Same as on the top plane, except  along the angular momentum component $Lz$ (bottom axis) and the corresponding radius of the circular orbit $R_G$ (top axis). Vertical dashed lines show the nominal $L_z$-values corresponding to the clusters detected in the velocity space. The black cross indicates the location of the Sun.}
    \label{fig:histogram}
\end{figure}

A common approach to studying the distribution of stars is to determine the stellar density as a function of the Galactocentric distance $R$. The resulting non-normalized stellar densities for the three samples as functions of $R$ are presented on the top panel in Fig.\,\ref{fig:histogram}. The Sun, which is a member of the thin disc, is located at the peak of the histogram corresponding to sample \textbf{\#1} (black cross). All curves exhibit a nearly symmetric distribution about the mean value, with varying peak amplitudes, but all centered at the Sun’s position.

\begin{figure}
    \includegraphics[width=1.\columnwidth]{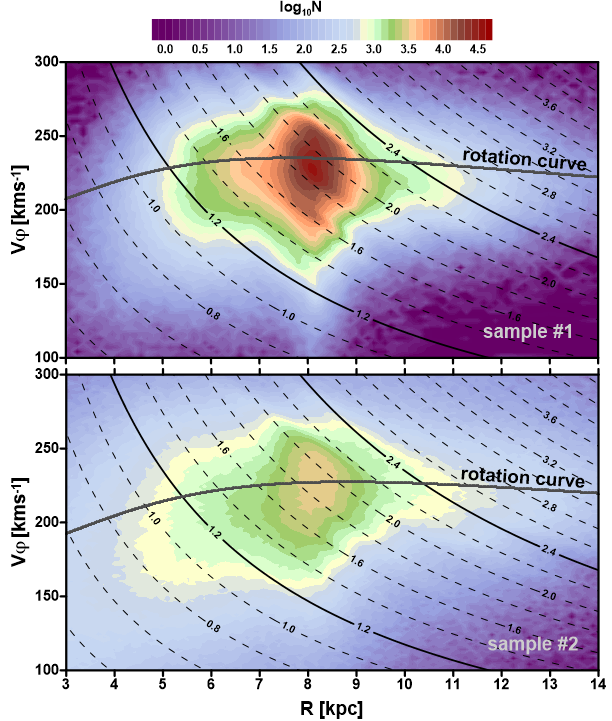}
    \includegraphics[width=1.\columnwidth]{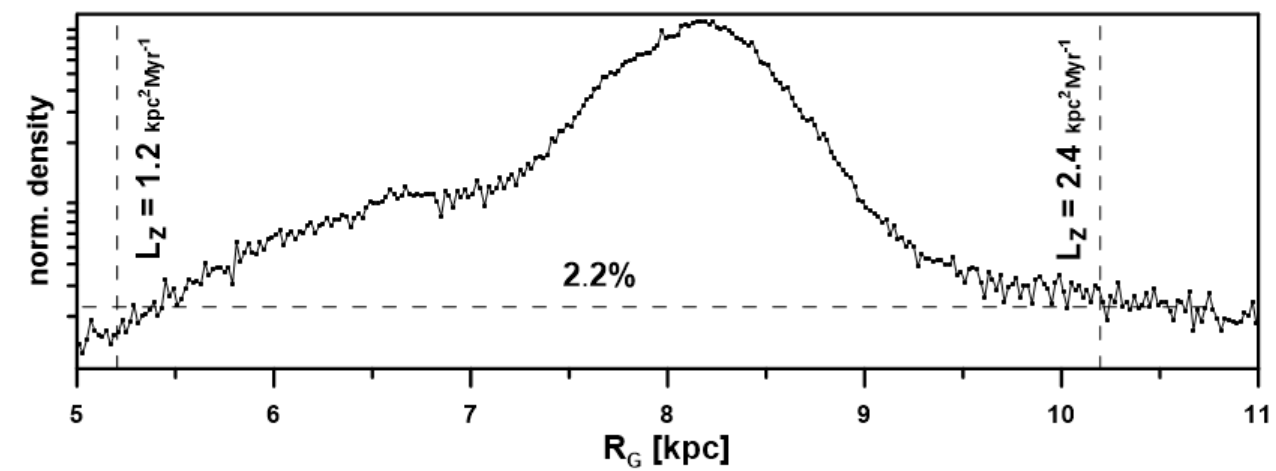}
    \caption{Distribution of stars from the sample \textbf{\#1} (top) and the sample \textbf{\#2} (middle) on the $R$--$V_\varphi$-plane; the bins of the maps are  0.1\,kpc per 2\,km\,s$^{-1}$. The color bar indicates the number of stars N (in logarithmic scale) in the bins. The dashed lines are levels of the angular momentum component $L_z=R V_\varphi$. The $L_z$-levels shown by continuous curves indicate the boundaries of the confidence zones on each panel. The location of the rotation curve is shown on both planes. 
    Bottom: the distribution of the nearly circular orbits from the sample \textbf{\#1} as a function of $R_G$. {The lower and upper limits of $R_G$ are indicated by the two vertical dashed lines at 5.2\,kpc and 10.2\,kpc, respectively.} }
    \label{fig:R-V-plane}
\end{figure}

In this study, we analyze the stellar density and kinematics using the angular momentum component $L_z$, a key quantity in stellar dynamics.
Indeed, since the perturbations from the spiral arms and the central bar are weak, stellar trajectories in the solar neighborhood follow the $L_z$-levels, indicated by dashed lines on the $R$--$V_\varphi$ planes in Fig.\,\ref{fig:R-V-plane}. The motion on the orbits sharing the same $L^\star_z$-value is an oscillation around a singular circular orbit of radius $R_G$ determined as a solution of the equation
$$
 RV_{\textrm{rot}}(R) - L^\star_z  = 0,
$$
where $V_{\textrm{rot}}(R)$ is the rotation curve\footnote{{The fitting procedure of the observed rotation curve is described in detail in \cite{2026ApJ..1000..236N}}.}, shown by continuous lines on the panels in Fig.\,\ref{fig:R-V-plane}. According to the definition, the radius $R_G$ of the circular orbit, also known as a ``guiding radius", can be uniquely associated with the corresponding value of $L_z$. For the Sun, $L_{z,\odot}=2.039$\,kpc$^2$\,Myr$^{-1}$, and the corresponding radius of the circular orbit is 8.5\,kpc.

To construct the density of stars as a function of $L_z$, we calculate the number of stars in a narrow bin around a given $L_z$-level and associate it with the $L_z$-value and the corresponding $R_G$. The result is shown on the bottom panel in Fig.\,\ref{fig:histogram}, for the samples from $\#1$ to $\#3$.
The upper horizontal axis of the graph shows the values of the radius $R_G$ corresponding to $L_z$ on the $x$-axis. The position of the Sun that belongs to the sample \textbf{\#1} is shown by the cross symbol at $L_z=2.039$\,kpc$^2$\,Myr$^{-1}$ and $R_G=8.5$\,kpc.
In contrast to the distribution along $R$ (top panel), we now observe clear corrugation patterns, indicating that the stellar population is not uniformly distributed as a function of $L_z$. This behavior may be related to stellar over-densities or agglomerations occurring at specific values of $L_z$. Cluster formations are more pronounced in sample \textbf{\#1} and closely resemble those observed in sample \textbf{\#3}; their nominal positions are indicated by  vertical dashed lines at specific  values of $L_z$.

The density profile of the thick disc (sample \textbf{\#2}), shown in the bottom panel of Fig.\,\ref{fig:histogram}, differs significantly from that of the thin disc (sample \textbf{\#1}), despite both samples containing a comparable number of stars. {This may indicate that even if stellar enhancements occur at specific values of $L_z$, they are too small or diffuse to be distinguished from the background noise in the thick disc sample.}
This observation accounts for the close similarity between the profile of the sample \textbf{\#3} and that of the “equatorial” sample \textbf{\#1}. Specifically, after normalization to their respective maxima and subsequent superposition, the two curves are nearly indistinguishable. Thus, in what follows, we restrict our analysis to the samples \textbf{\#1} and  \textbf{\#2} only.

\subsection{Analysis of the confidence region}\label{sec:sub-sec-1}

Next, we define a confidence region in terms of both $L_z$ and $R_G$ and, within this region, {we consider our sample to be sufficiently complete and representative to ensure the reliability and scientific validity of the conclusions drawn from the analysis.}

For this task, we plot on the representative planes $R$--$V_\varphi$ in Fig.\,\ref{fig:R-V-plane} the densities of stars from the samples \textbf{\#1} (top panel) and \textbf{\#2} (bottom panel). The number of stars in bins of 0.1 \, kpc per 2 \, km \, s$^{-1}$ has a maximal value at 20,000 in the thin disc and 3,000 in the thick disc. 
The color bar on the top (same for both panels) relates the density values (in logarithmic scale) to the corresponding range of colors.
Two solid $L_z$-levels enclose all map bins that count for at least $10^3$ (an arbitrary threshold) stars for both samples on the panels, between  a lower value at 1.2\,kpc$^2$\,Myr$^{-1}$ and an upper value at 2.4\,kpc$^2$\,Myr$^{-1}$, or, in terms of $R_G$, 5.2\,kpc and 10.2\,kpc, respectively.

{The derived bounds on $R_G$ can be confirmed by examining how the circular orbits are distributed along $R$. Indeed}, by definition, $R_G$ denotes the radius of the circular orbits that are on the rotation curve. Following this definition, we select from the sample \textbf{\#1} all stars that lie in the close vicinity of the rotational curve, with velocities satisfying the conditions $|V_R|<5$\,km\,s$^{-1}$ and $|V_\varphi - V_{\textrm{rot}}(R)| < 4$\,km\,s$^{-1}$. The total number of nearly circular orbits in the sample \textbf{\#1} is around 116000 stars. For these stars, whose orbits satisfy $R \approx R_G$, the number distribution as a function of $R_G$ is presented (in logarithmic scale) on the bottom panel in Fig.\,\ref{fig:R-V-plane}. {Two vertical dashed lines on the bottom panel in Fig.\,\ref{fig:R-V-plane} delimit the confidence zone, while the horizontal dashed line indicates that only a small fraction of the sample (2.2\%) is excluded from the analysis.}

\begin{figure}
    \includegraphics[width=1.\columnwidth]{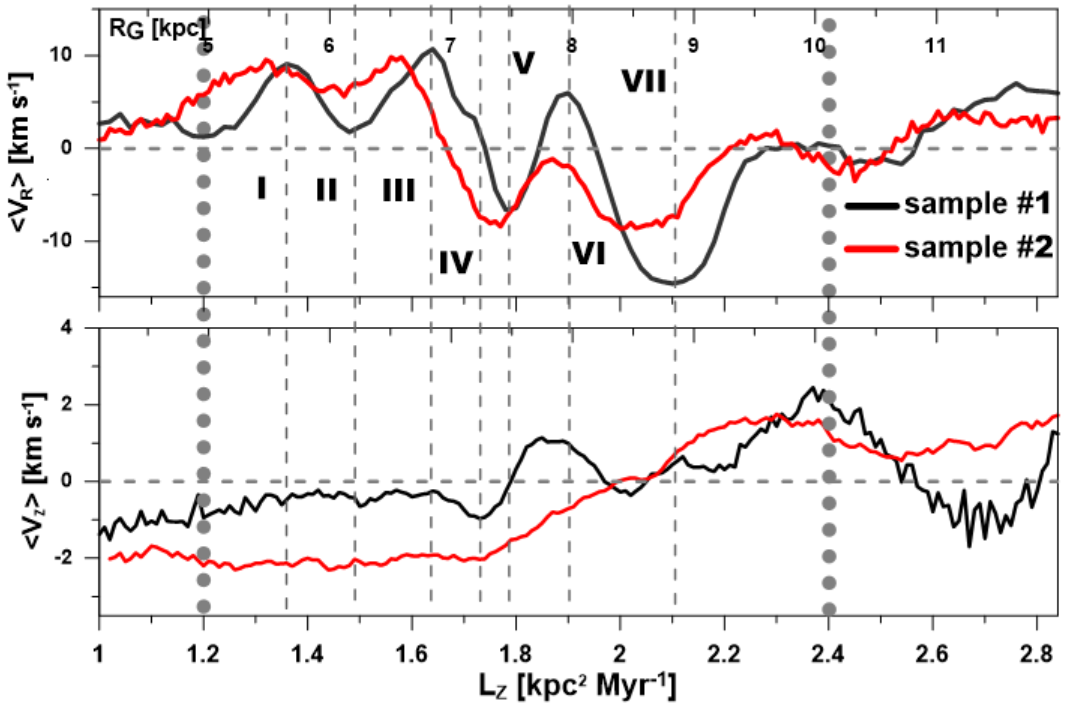}
    \caption{The radial (top) and vertical (bottom) velocity waves as functions of the angular momentum $L_z$ (bottom axes) and the radius  of the corresponding circular orbit $R_G$ (top axes). Black (red) curves correspond to the thin (thick) disc.
    The vertical dashed lines (denoted from \textbf{I} to \textbf{VII}) match the main peaks/valleys of the $\left<V_R\right>$-wave; their locations are listed in Table\,\ref{tab:waves}. The confidence zone is delimited by two vertical dotted lines. 
}
    \label{fig:fig-waves}
\end{figure}

\section{Velocity waves and corrugations along $L_z$}
\label{sec:waves}

Figure\,\ref{fig:fig-waves} shows the stellar radial (top) and vertical (bottom) velocity distributions, represented by their median values computed in narrow bins of $L_z$.
The radii of circular orbits $R_G$, corresponding to the $L_z$--values on the bottom axes, are shown on the top axes of the graphs. The confidence zone  is limited by the two vertical dotted lines, between 1.2 and  2.4 (in units kpc$^2$\,Myr$^{-1}$). {All curves display increased noise levels outside this interval, where low counting statistics prevail as a result of sample incompleteness.}

The observed oscillatory patterns and the corrugations in the velocity distributions are already known in the literature as ``velocity waves". 
A comparison of the  $\left<V_R\right>$-waves in the thin and thick discs on the top panel in Fig.\,\ref{fig:fig-waves} shows  qualitatively similar behavior. The $\left<V_R\right>$ oscillations of thick disc stars (red curve in the top panel) track those of the thin disc, but with reduced amplitudes and a slight offset. This indicates that the processes responsible for ripples in the radial component of stellar velocities continue to operate outside the equatorial Galactic disc.

We identify the locations of the peaks and valleys of the $\left<V_R\right>$-wave in the sample \textbf{\#1}  by vertical dashed lines and symbols from \textbf{I} to \textbf{VII} in Fig.\,\ref{fig:fig-waves}; their coordinates $L_z$ and $R_G$ are listed in Table\,\ref{tab:waves}. It should be noted that the features labeled \textbf{III} to \textbf{VII} exhibit the same coordinates as the stellar {enhancements} identified in  Fig.\,\ref{fig:histogram} (bottom). This coincidence suggests that they are likely structures of the same origin.

\begin{table}
\caption{Extrema of the $\left<V_R\right>$-wave of sample \textbf{\#1}.\\
\small{units: $L_z$ [kpc$^2$\,Myr$^{-1}$], $R_G$} [kpc]}
\label{tab:waves}      
\centering   
\begin{tabular}{l c c c }
\hline\hline 
Symbol  & $L_z$ & $R_G$ & MG \\
[0.5ex]
\hline
\textbf{I} & 1.36 & 5.75 & HR1614 \\
\textbf{II} & 1.49 & 6.25 & Her-1 \\
\textbf{III} & 1.64 & 6.84 & Her-2 \\
\textbf{IV} & 1.73 & 7.20 & Her-3 \\
\textbf{V} & 1.79 & 7.45 & Pleiades \\
\textbf{VI} & 1.90& 7.90 & Hyades \\
\textbf{VII} & 2.10 & 8.77 & Coma-Sirius \\
[0.5ex]  
\hline\hline
\\
\end{tabular}
\end{table}

\begin{figure}
     \includegraphics[width=1.0\columnwidth]{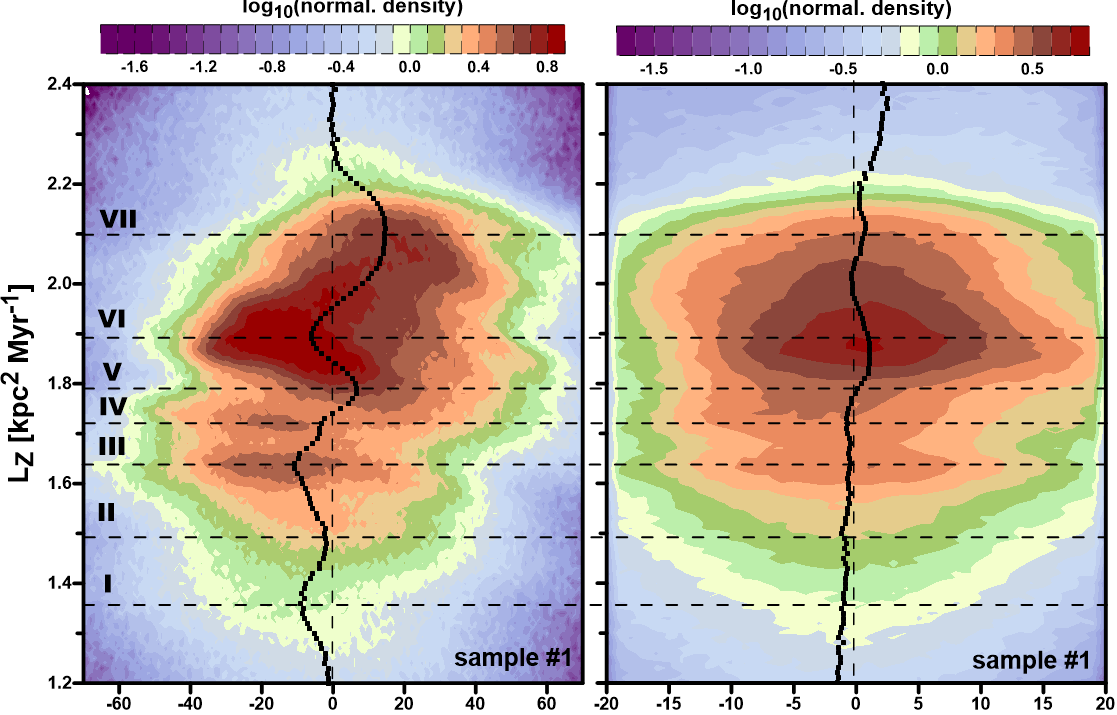}
     \includegraphics[width=1.0\columnwidth]{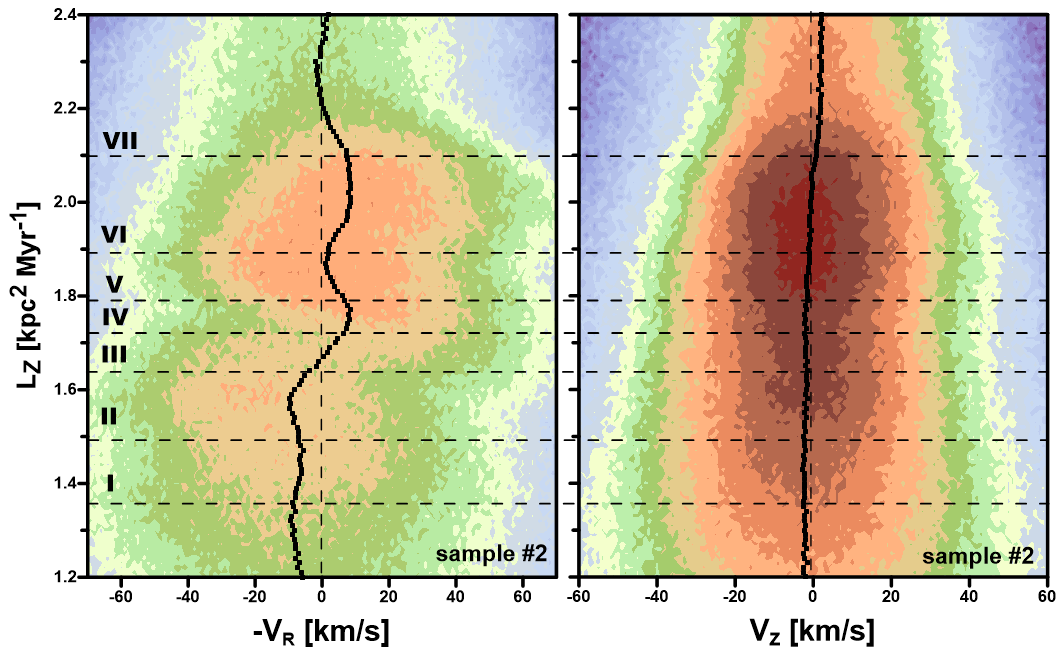}
    \caption{The normalized stellar density mapped by the color scale on the $-V_r$--$L_z$-planes (left panels) and the $V_z$--$L_z$ planes (right panels). The thin (thick) disc stars are shown on the top panels (bottom panels). The $\left<V_R\right>$- and $\left<V_z\right>$-waves are superimposed on the  corresponding planes. The horizontal dashed lines from \textbf{I} to \textbf{VII} indicate the extrema values of the $\left<V_R\right>$-wave of the sample\,\textbf{\#1} and the corresponding nominal $L_z$-locations of the main MGs.}
    \label{fig:fig-sample1}
\end{figure}

The vertical velocities shown on the bottom panel in Fig.\,\ref{fig:fig-waves} exhibit behavior that differs from the  noticeable ripples observed in the radial velocities. Some picks/valleys of small amplitudes (up to 2\,km\,s$^{-1}$) are still observed in $\left<V_z\right>$ in thin disc (black curve), but their relationship to  radial velocity waves  is not clear. Most likely, those that are located in the interval between extrema \textbf{III} and \textbf{VII}  of the $\left<V_R\right>$-wave, between 1.74 and 2.25 (in units kpc$^2$\,Myr$^{-1}$), are related to the main MGs, Pleiades, Hyades, and Coma-Sirius. In contrast, for thick-disc stars in the sample \textbf{\#2}, the mean vertical velocity $\left<V_z\right>$ rises steadily with $L_z$ within this interval and remains constant beyond it. Understanding this pattern in the vertical velocity of stars in the thick disc will require more in-depth studies.

\section{Velocity waves and the main moving groups} 
\label{sec:moving groups}

The oscillatory behavior of the median velocities as a function of the angular momentum component $L_z$ can be understood by analyzing the velocity distributions within individual $L_z$ bins.
For stars from thin and thick disks, whose $L_z$-values lie within the confidence interval ($1.2<L_z<2.4$, in units of kpc$^2$\,Myr$^{-1}$), we construct maps of the representative planes $V_R$--$L_z$ and $V_z$--$L_z$.  The maps shown in Fig.\,\ref{fig:fig-sample1} depict stellar density (normalized to the average density over the domain), using a color contour map. 
The $\left<V_R\right>$- and $\left<V_z\right>$-waves discussed in the previous section are superimposed on the $V_R$–$L_z$ and $V_z$–$L_z$ planes, respectively. The horizontal dashed lines mark the $L_z$-positions of the $V_R$-extrema listed in Table\,\ref{tab:waves}.

The asymmetric pattern in the distribution of stars seen on the $V_R$–$L_z$ velocity plane of the thin disc (top-left panel) has already been studied in relation to the phenomenon of \textit{moving groups} (MGs)\footnote{To remain consistent with the conventional representation of stellar moving groups, we adopt the radial velocity component ($V_R$) with reversed sign.} (e.g., \citealt{Dehnen2000, Michtchenko2018ApJL, HUNT2025NEWAR}). 
By analyzing the median value of $-V_R$ along the dashed horizontal lines, we can understand how the maximum and minimum values of $\left<V_R\right>$ vary as functions of $L_z$: these extrema are directly connected to the known locations of the main MGs. We list in Table\,\ref{tab:waves} the main MGs, which are associated with the min/max of the $\left<V_R\right>$-wave: they are \textbf{I} - HR1614, \textbf{II}-\textbf{IV} - three branches of Hercules, \textbf{V} - Pleiades, \textbf{VI} - Hyades, \textbf{VII} - Coma Berenices  and Sirius.

Using the same color scale, we analyze the asymmetries in the stellar distribution  of the thick disc shown on the bottom-left panel in Fig.\,\ref{fig:fig-sample1}. In this case, the stellar density decreases by about an order of magnitude (see Figs.\,\ref{fig:histogram} and \ref{fig:R-V-plane}), but signatures of stellar clustering still remain visible, enabling us to trace the contours of the MGs. 
As in the case of the thin disc, the connection between the $\left<V_R\right>$-extrema and the MGs can be observed. {The small offsets of the peaks of the $\left<V_R\right>$-wave from the nominal $L_z$-positions of the groups listed in Table\,\ref{tab:waves} may arise from the coupling between the radial and vertical modes of stellar motion (e.g., \citealp{2023Michtchenko}). However, a detailed investigation of this effect is beyond the scope of the present study.}

The $V_z$-$L_z$ maps presented in Fig.\,\ref{fig:fig-sample1} do not clearly show the asymmetric patterns, which are characteristic of the $V_R$-$L_z$ maps. The $\left<V_z\right>$-wave curves overlaid on the plots enable us to identify the asymmetries exclusively in the thin disc (top-right panel): they are located between the \textbf{IV} and \textbf{VII} groups, Her-3 and Coma/Sirius, respectively, and {their amplitudes are small; the maximum detected $\left<V_z\right>$-amplitude for group \textbf{IV} is approximately 1\,km\,s$^{-1}$.} The behavior of the $V_z$-velocity as a function of $L_z$ of the thick disc on the bottom-right panel shows no notable features.

\begin{figure}
    \includegraphics[width=1.0\columnwidth]{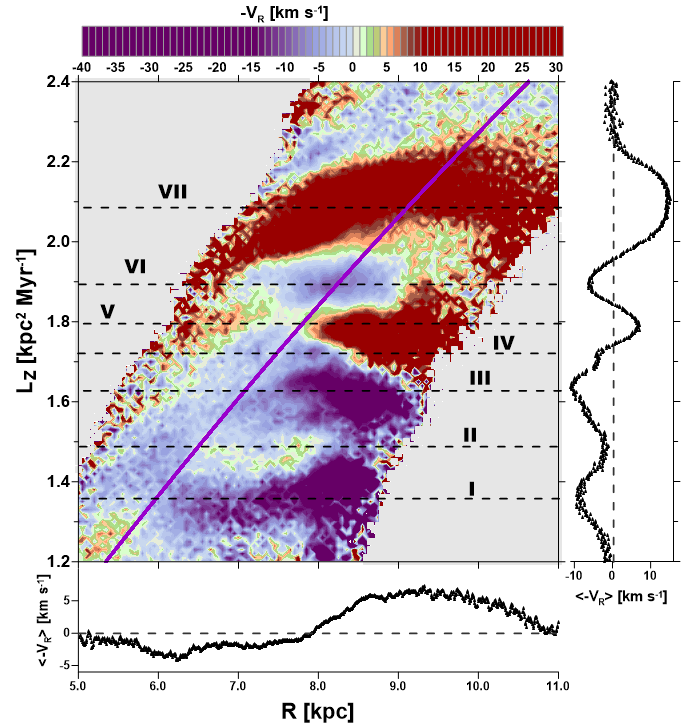}
    \caption{The main moving groups from \textbf{I} to \textbf{VII} on the $R$--$L_z$ plane, using the median value of $V_R$ as an indicator; when plotted on the  $R$--$V_\varphi$ plane, they are known as \textit{diagonal ridges}. The solid line is the rotation curve obtained for the sample \textbf{\#1}. The horizontal dashed lines from \textbf{I} to \textbf{VII} correspond to the extrema of the $\left<V_R\right>$-wave of sample\,\textbf{\#1}, and the position of the well-known moving groups on the radial velocity-action plane. }
    \label{fig:fig-fig5}
\end{figure}

\begin{figure*}
    \includegraphics[width=2.\columnwidth]{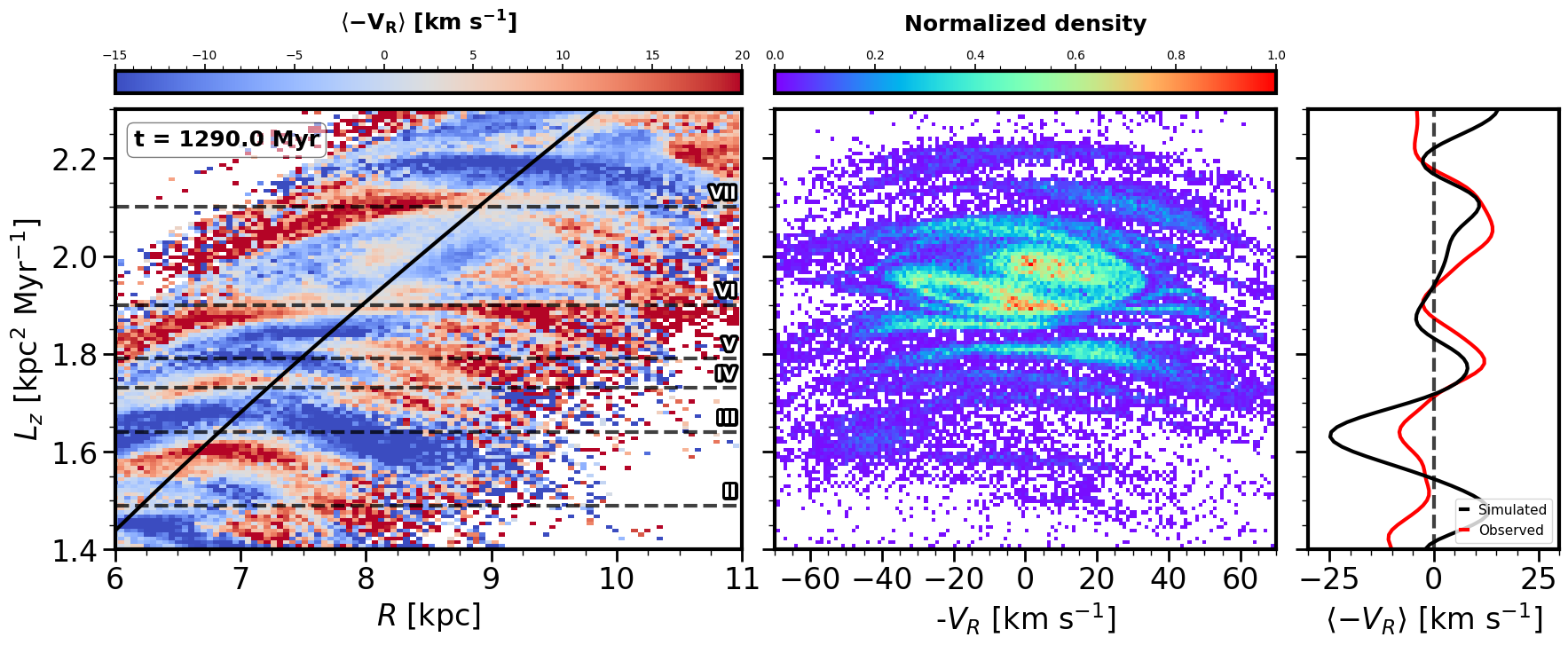}
    \caption{Distributions of the test particles on the $R$--$L_z$ plane (left panel) and the $V_R$--$L_z$ plane (middle panel), resulting from the simulation, at $t=1.29$\,Gyr. On the left panel, the solid line represents the rotation curve derived for the thin disc. The horizontal dashed lines, labeled \textbf{II} through \textbf{VII}, denote the $L_z$ values associated with the corresponding MGs reported in Table\,\ref{tab:waves}.
    Right: comparison between the simulated (black) and observed (red) $V_R$-velocity waves. Note that both curves were calculated for the stars within the 200\,pc from the Sun.
    }
    \label{fig:fig-simulation}
\end{figure*}

\section{Velocity corrugations along $R$}\label{sec:ridges}

The spatial extension of the MGs as a function of the Galactocentric radius $R$ is typically investigated on the $R$--$V_\varphi$ phase-space plane, where the MGs manifest themselves as characteristic “diagonal ridges” (DRs) (e.g., \citealp{2020ApJ...888...75B, 2021FrASS...8...48B, 2026ApJ..1000..236N}). To investigate how the $\left<V_R\right>$-waves are connected to the DRs, in Fig.\,\ref{fig:fig-fig5} we present, with a color scale, the distribution of the median $-V_R$ for thin disc stars on the $R$--$L_z$ plane. 
As indicated by the color bar, the blue areas indicate negative values of $\left<-V_R\right>$, while the red areas indicate positive values of $\left<-V_R\right>$.
The solid line denotes the rotation curve of the thin disc, corresponding to the loci of circular orbits for which $V_R=0$. 
The horizontal dashed lines indicate the $L_z$-positions of the extrema of  $\left<V_R\right>$ listed in Table\,\ref{tab:waves}.
The bottom plot shows the $R$--$\left<-V_R\right>$-plane, while the right plot shows the $\left<-V_R\right>$--$L_z$-plane.

The DRs are visible on the main panel in Fig.\,\ref{fig:fig-fig5} as features of various sizes and colors, all extending  along the constant levels of $L_z$ listed in Table\,\ref{tab:waves}. The relationship between the values of $-V_R$ averaged over the levels of $L_z$ and the extreme values of $\left<-V_R\right>$ shown in the right plot in the main panel can be easily identified. 

In contrast, studying the behavior of the radial velocity as a function of $R$, the averaging is done over the vertical constant $R$-levels on the same plane. The resulting function is plotted on the bottom plot in Fig.\,\ref{fig:fig-fig5}. It is clear that, for $R<8$\,kpc, negative values of $-V_R$ (blue tones) prevail on the $R$--$L_z$ plane, while for $R>8$\,kpc the opposite pattern is observed.

Thus, the apparently different patterns of the averaged $V_R$ as functions of $R$ and $L_z$ arise solely due to the distinct projections of the same structures, which appear as MGs in the velocity plane, but as DRs in the space–action plane. In this way, we have clarified the question raised in \citet[]{HUNT2025NEWAR}, namely, ``whether the pattern in $R$ is simply a smooth representation of the pattern in $R_G$ caused by `blurring’ due to stars’ epicyclic motions, or whether they constitute two separate dynamical signals."

\section{Simulation of the velocity waves} \label{sec:simulations}

To corroborate our findings derived from the analysis of the spatial and kinematic distributions of stars, we carried out some numerical simulations of stellar motions in the Galactic disc.  Following \citet{2020ApJ...888...75B}, we performed test-particle integrations on a 3\,Gyr, during which we tracked the evolution of the stellar radial velocity behavior.
Focusing on the dynamics of the stars in the thin disc, we allowed the stars to evolve under forces of the 2D Galactic axisymmetric potential and combined perturbations from the spiral arms and the central bar (\citealp[]{2026ApJ..1000..236N}). 

We assumed that star formation occurs predominantly in the spiral arms due to the higher density of gas and dust. Thus, the initial conditions of $\sim 1.4\times10^8$ stars were randomly placed across the four spiral arms, the Local Arm and the background disc, according to their respective probability distributions.  
The initial radial velocities $V_R$ of the star particles were obtained from a Gaussian distribution with zero mean, whereas the initial tangential velocities $V_\varphi$ were randomly sampled from a Gaussian distribution centered on the rotation curve, both using dispersions that depended on the Galactocentric radius. For a detailed justification for the initial conditions adopted, the reader is referred to Sect.\,3 in \cite{2020ApJ...888...75B}.

By construction, the initial velocity field is symmetric. However, this symmetry is rapidly broken during the earliest stages of the evolution under the influence of the non-axisymmetric gravitational perturbations associated with the central bar and the spiral arms. The phase-mixing stage, during which density structures emerge and continuously alter their shapes and locations on the plane, persists for roughly 1\,Gyr, after which the process stabilizes. 

{Figure \ref{fig:fig-simulation} shows the simulated distributions of the median radial velocity (left panel), the density distributions of the test particles (middle panel) and the simulated $\left<V_R\right>$--wave (right panel) at instant $t = 1.29$\,Gyr. The nominal locations of the main MGs listed in Table\,\ref{tab:waves} are indicated by horizontal dashed lines and the corresponding numbers on the $R$--$L_z$ plane, which exhibits clear similarities with those reported in Fig.\,\ref{fig:fig-fig5}. The main structures that appear on the simulated $V_R$--$L_z$ plane include arches, streams and edges, among others, which were previously observed by \cite{2018Natur.561..360A} and interpreted in the framework of spiral-arm–induced perturbations  by    \cite{Michtchenko2018ApJL}.}

The simulated $\left<V_R\right>$-wave (black line) is compared to the observed $V_R$-velocity wave (red curve) on the right panel in Fig.\,\ref{fig:fig-simulation}. The good agreement between the curves supports our conclusion that the asymmetric patterns of the radial velocity distribution arise from internal dynamical processes. The only notable discrepancy is identified in the vicinity of the lower boundary of the confidence $L_z$-zone, where the stellar density is comparatively low.

\section{Summary and Discussions} 
\label{sec:conclus}

The kinematic properties of stars in the \textit{Gaia DR3} catalog are analyzed by characterizing the distribution of their orbital parameters in phase space.  When presented as functions of $L_z$, both the density profile and the median radial and vertical velocities display oscillatory patterns and corrugation, with the effect being most prominent for stars in the thin disc.  By cross-matching the local extrema of the $V_R$-wave with the corresponding structures identified on both the $V_R$–$L_z$ and $R$–$L_z$ planes, we established a robust association between the radial velocity wave and the principal MGs and the DRs. 

The choice of the angular momentum component \(L_z\) as the tracer was a crucial aspect of this study, and this choice is based on a specific reason. Indeed, as shown in Sect.\,\ref{sec:sample}, $L_z$ is a key quantity in stellar dynamics, since it is a constant of motion in the unperturbed model of the  axisymmetric Galactic disc. Even when the perturbative contributions of the spiral arms and the central bar are incorporated into the Galactic potential, their effects remain several orders of magnitude smaller than the underlying axisymmetric component. Consequently, stellar trajectories generally continue to evolve approximately along surfaces of constant $L_z$. However, there exist zones in the phase space where the resonant  coupling between the radial and azimuthal components of stellar motion provokes large amplitude $L_z$-oscillations. These zones form the chains of resonant islands of stability, and  accumulation of stars inside them. This phenomenon is commonly referred to as corotation (CR) and Lindblad resonances (LR), which constitute a mechanism for the formation of MGs in the velocity subspace of the phase space and DRs in the configuration subspace (e.g., \citealp[]{2020ApJ...888...75B}; \citealp{Michtchenko2018ApJL}).

If the perturbations persist over a sufficiently long timescale, the dynamical system evolves toward a resonant equilibrium state, characterized by a complex and structured phase-space portrait. 
The ability of the resonant model to reproduce the structures observed in the stellar kinematic distributions indicates that the local Galactic region is currently in a state of dynamical equilibrium. Our numerical simulations of the evolution of the stellar distribution in the solar region, from an initially homogeneous configuration to a resonantly structured state, have shown that the system attains dynamical stability over a timescale of approximately 1\,Gyr.  {The detailed characterization of the perturbation effects driven by the Galactic central bar and the spiral arms reported by \cite{2026ApJ..1000..236N} enabled us to associate CRs and LRs with the known MGs, the DRs and the $V_R$-wave on the parametric planes $\Omega_\textrm{bar}-L_z$ and $\Omega_\textrm{sp}-L_z$, where $\Omega_\textrm{bar}$ and $\Omega_\textrm{sp}$ are pattern velocities of the central bar and the spiral arms, respectively (see Figure\,10 of that paper).}

We further note that the wave-like patterns in the median velocities progressively diminish in amplitude with increasing altitude toward the thick disc region. Notably, the waves associated with the corotation zone, which includes the Pleiades, Hyades and Coma-Sirius groups, 
exhibit a more pronounced vertical extension, whereas those located at other resonance radii rapidly dissipate from the Galactic mid-plane (Fig.\,\ref{fig:fig-waves}). Nevertheless, the vertical stellar distribution reveals intriguing signatures that require further investigation. These features include the pronounced plateau in the stellar density profile shown by the red curve in the bottom panel of Fig.\,\ref{fig:histogram}, as well as the monotonic increase with Galactocentric distance $R$ and the associated shifts in the mean vertical velocity evident in Fig.\,\ref{fig:fig-waves}. It should be noted that, in contrast to extensive investigations of resonant phenomena within the Galactic equatorial plane, analyses of the corresponding dynamics in the vertical direction remain comparatively scarce (\citealp{2020ApJ...890...85L}; \citealp{2022Bernet}; \citealp{2023Michtchenko}).

Finally, our findings indicate that the oscillatory behavior of the median radial and azimuthal velocities must be considered when using them as tracers of the vertical stellar distribution. For instance, in \cite{Michtchenko2019ApJ} we demonstrate that the vertical phase-space spiral in the $Z$--$V_z$ subspace is generated by the MGs, mainly by Coma-Berenices, Hyades \& Pleiades, and Sirius, when the statistical characteristics (mean, median, or mode) of the radial and azimuthal velocities are used to analyze the distribution in the vertical position-velocity plane. Our conclusion has received partial observational corroboration in the recent work by \cite{2026ApJXu}.

\section*{Acknowledgements}

The authors thank our anonymous reviewer for his helpful comments and suggestions. 
The authors also thank Dr. Phillip Galli for useful suggestions concerning the analysis of Gaia’s astrometry data.
This work was supported by the Brazilian funding agencies CNPq, FAPESP, and CAPES.  W.S.D. is Bolsista FAPEMIG—CNPq—Brasil (processo BPQ-06608-24). This work has made use of data from the European Space Agency (ESA) mission Gaia (\url{https://www.cosmos.esa.int/gaia}), processed by Gaia Data Processing and Analysis Consortium (DPAC, \url{https://www.cosmos.esa.int/web/gaia/dpac/consortium}). This work has made use of the facilities of the Laboratory of Astroinformatics (IAG/USP, NAT/Unicsul), funded by FAPESP (grant 2009/54006-4) and INCT-A.


\bibliographystyle{apj}
\bibliography{references-1}




\end{document}